\documentclass[namedreferences]{SolarPhysics}
\usepackage[optionalrh,solaenum]{spr-sola-addons} 
\usepackage{graphicx}                    
\usepackage[usenames]{color}                       
\usepackage{url}                         
\usepackage{color}

\newcommand{\aap}{    {\it Astron. Astrophys.}}

\newcommand{\apj}{    {\it Astrophys. J.}}
\newcommand{\apjl}{   {\it Astrophys. J. Lett.}}

\newcommand{\jgr}{    {\it J. Geophys. Res.}}

\newcommand{\solphys}{{\it Solar Phys.}}

\begin{document}

\begin{article}

\begin{opening}

\title{Interchange Slip-Running Reconnection and Sweeping SEP Beams}

\author{S.~Masson{}$^{1}$\sep G.~Aulanier{} $^{2}$\sep E.~Pariat{} $^{2}$\sep K.-L.~Klein{} $^{2}$\sep}
%

%
   
 \institute{$^{1}$ Space Weather Laboratory, NASA Goddard Space Flight Center, 8800 Greenbelt Road, Greenbelt, MD 20771, USA
 email : \url {sophie.masson@nasa.gov}\\
 $^{2}$LESIA, Observatoire de Paris, CNRS, UPMC, Universit\'e 
                     Paris Diderot, 5 place Jules Janssen, 92190 Meudon, 
                     France,
         		 }
\begin{abstract}

 We present a new model to explain how particles (solar energetic particles; SEPs), accelerated at a reconnection site that is not magnetically connected to the Earth, could eventually propagate along the well-connected open flux tube. Our model is based on the results of a low-$\beta$ resistive magnetohydrodynamics simulation of a three-dimensional line-tied and initially current-free bipole, that is embedded in a non-uniform open potential field. The topology of this configuration is that of an asymmetric coronal null-point, with a closed fan surface and an open outer spine. When driven by slow photospheric shearing motions, field lines, initially fully anchored below the fan dome, reconnect at the null point, and jump to the open magnetic domain. This is the standard interchange mode as sketched and calculated in 2D. The key result in 3D is that, reconnected open field lines located in the vicinity of the outer spine, keep reconnecting continuously, across an open quasi-separatrix layer, as previously identified for non-open-null-point reconnection. The apparent slipping motion of these field lines leads to form an extended narrow magnetic flux tube at high altitude. Because of the slip-running reconnection, we conjecture that if energetic particles would be traveling through, or be accelerated inside, the diffusion region, they would be successively injected along continuously reconnecting field lines that are connected farther and farther from the spine. At the scale of the full Sun, owing to the super-radial expansion of field lines below $3~\rm{R_{\odot}}$, such energetic particles could easily be injected in field lines slipping over significant distances, and could eventually reach the distant flux tube that is well-connected to the Earth.

\end{abstract}

\end{opening}

\section{Introduction}
\label{intro} 

 The coronal magnetic field is the main driver of the eruptive solar activity and ensures the coupling between the corona and the interplanetary medium. One of the manifestation of the solar activity is the acceleration  of solar energetic particles (SEPs), electrons and ions, at energy ranges from few keV to few GeV and their release in the heliosphere. In the Sun's corona, these energetic particles can be accelerated either at low altitudes inside magnetic reconnection regions, {\it i.e.} in flare locations, or at higher altitudes at the shock driven by coronal mass ejections (CMEs).

It is generally believed that cross-field diffusion is weak during the interplanetary transport of energetic particles \cite{Droege10}. According to that, energetic particles can only reach the Earth if they are eventually injected along the very field lines connecting the acceleration site and the Earth. Hereafter, we refer to the latter as ``Earth-connected field lines''. The Solar Particles Events are divided in two categories that depends on the time profile of energetic particle flux measured at Earth : the gradual-SEP and impulsive-SEP. The gradual events are usually considered as being accelerated through the CME-driven shock, implying a large longitudinal injection of energetic particles \cite{Reames99}.  Thus, energetic particles eventually reach the ``Earth-connected field lines''  \cite{Cliver_al04,Nitta_al06}. On the contrary, the impulsive events are accelerated through the flare \cite{Cane_al86,Reames99} and their interplanetary injection may not be straightforward. Indeed, according to several studies \cite{Lin70,Kallenrode_al92,Nitta_al06}, energetic protons and electrons impacting the Earth may be related to flares located far from the footprints of the  Earth-connected field line.  Moreover, it is noteworthy that the flare-accelerated particles are sometimes detected by different spacecraft  in the heliosphere over a wide longitudinal range around Earth-connected field lines \cite{ReinWibb74,WibbCane06}. 

These flare-accelerated particles are believed to be accelerated at low altitude within the corona and not at the CME-driven shock. Thereby, their injection into the interplanetary medium should be achieved through a reconnection process between open and closed coronal magnetic field. This reconnection can either be the same as, or different from, the flare reconnection.

 This paper addresses the following questions : How can magnetic reconnection allow energetic particles that are accelerated in the low solar corona, far from the ``Earth-connected field lines'', to get access to the Earth ? Do exist dynamics of the magnetic reconnection that would lead to inject flare-accelerated particles over  a wide angular distance in the heliosphere ? Hereafter, we link the three issues of the open versus closed corona, of particle injection problem and of three-dimensional reconnection, in order to motivate the model that we propose to answer these two questions.

\subsection{The Coupling Between the Closed Corona and the Heliosphere}

The large-scale coronal magnetic field presents two different  types of connectivity: closed and open. In the closed magnetic domain, field lines have both footpoints rooted in the photosphere, while for the open domain, a field line is anchored at the photosphere by one footpoint only, its other end extending into the heliosphere. The coupling between the corona and the interplanetary medium is carried out at the transition between open and closed magnetic field. This coupling can either be ideal ({\it e.g.} when propagating waves carry energy from one domain to another) or non-ideal ({\it e.g.} when magnetic reconnection leads to mass and magnetic flux transfer between both domains). Examples of coronal magnetic configurations displaying such an open/closed transition are :  at large-scales, the helmet streamers and plumes/pseudo-streamers ({\it e.g.}  \opencite{Wang_al07a}, \opencite{Wang_al07b});  at smaller scales, coronal jets ({\it e.g.} \opencite{Cirtain_al07}). Streamers are formed above bipolar regions, where the closed coronal loop system is opened at high altitudes to the interplanetary medium through the solar wind pressure ({\it e.g.} \opencite{Grappin_al02}). Pseudo-streamers, and jets respectively, are created when a larger active region ({\it see e.g.} \opencite{DelZanna_al11} ) or a small magnetic bipole \cite{Shibata_al94,Moreno_al08,Torok_al09} emerges in an already open coronal magnetic flux tube.

It has been shown that, apart from special cases \cite{Titov_al10}, a single bipole embedded in an open field most often relates to a magnetic topology that comprises a single magnetic null point in the corona ({\it as in} \opencite{Antiochos98}; \opencite{Pariat_al09}; \opencite{Torok_al09}). Such a null-point is associated to a fan separatrix surface and a singular spine field line \cite{LauFin90}. For a bipole embedded in an open field, the fan forms a dome which encloses an inner connectivity domain. The fan separatrix surface separates closed magnetic field lines from the surrounding open ones, located in the outer connectivity domain. The spine field line crosses the fan surface at the null point. Therefore, its outer part is also open. 

Magnetic reconnection at the null point allows the exchange of connectivity between closed and open magnetic field lines. This is the standard ``interchange reconnection mode'' \cite{Pariat_al09,Edmondson_al09}. In general, this closed to open reconnection allows a direct coupling between the open and the closed corona. In particular, it is a natural mechanism to inject impulsive-SEP related energetic particles from the closed corona to the interplanetary medium. The acceleration may occur either in a flare located inside the inner domain, before the null point reconnection itself, or  at the null point during the interchange reconnection \cite{Litvinenko06,RosGals10,Browning_al10}. The detection at the Earth of such flare-accelerated particles implies that the open coronal flux tube surrounding the outer spine and involved in the interchange reconnection, must be {\em a priori} directly connected to the Earth-connected field lines. But, it is not always the case.

\subsection{The Problem of the Particle Injection}

Generally, the geometry of the interplanetary magnetic field (IMF) is assumed to follow the Parker spiral. Depending on the solar wind velocity, this implies that the Earth-connected field lines are rooted between $40^{\circ}$ and $70^{\circ}$ in longitude. However, past studies of impulsive-SEP events show that energetic particles can very often be detected at the Earth even if the associated flare is located away (up to $50^{\circ}$) from the Earth-connected interplanetary spiral field lines \cite{Kallenrode_al92,Nitta_al06}. This may first imply that the Earth-connected field line is not a simple Parker spiral starting in the low corona.

 Indeed, using PFSS extrapolations  \inlinecite{Klein_al08} showed that open magnetic flux tubes in flaring active regions associated to SEP events, can spread in longitude from $4^{\circ}$ at the photosphere to  more than $70^{\circ}$ at the solar source-surface, at the base of the Parker spiral. Comparing coronal magnetic field, metric radio observations and energetic particles in-situ measurements, they showed that the large longitudinal extensions of open magnetic flux tubes could ensure a direct path to Earth for energetic particles, being accelerated far from the Earth-connected field lines. Even though this effect is surely important, it may not be the only way to inject energetic particles onto the Earth-connected field lines.

In addition to this first issue, energetic particles of some impulsive SEP events can also be detected in a wide flux tube, over more than $90^{\circ}$ of longitude from the the parent active region \cite{WibbCane06}. These large particle beams can be explained by lateral transport along coronal magnetic loops \cite{NewkirkWentzel78} or lateral diffusion in open coronal magnetic field \cite{Reid64,Wibberenz_al89}. Instead invoking the transport effect of particles, could this large angular spread result from a specific dynamics of injection channel related to the interchange reconnection mode?

For conventional models of interchange reconnection, we expect that particles can only be injected into a narrow bundle of field lines whose connection to the Earth is rather unlikely.  Indeed, the field line that reconnects at the null point and opens in the outer connectivity domain is located where the outer spine was located at the time preceding the reconnection. Thus, as the reconnection evolves, the reconnected field lines form an open flux tube concentrated around the initial location of the outer spine.  Flare-accelerated particles should therefore be only injected into this localized flux tube. Even though the open magnetic field can display a rapid (super-radial) divergence in the corona (as easily seen  in \opencite{Wang_al07a}; \opencite{Klein_al08}), particles are transferred into the open corona along  this localized flux tube, implying a narrow localized particle beam. Nevertheless, the spine itself can shift its position by several degrees during a long-enough reconnection process  ({\it see eg.} \opencite{Moreno_al08}; \opencite{Edmondson_al09}), and could lead to broaden the open flux tube along which particles would be injected. But, we ignore if this motion is directed towards the Earth-connected field line and if the displacement covers a sufficient angular distance. This connectivity problem is illustrated by the theoretical geometry sketched in Figure~\ref{f-fig1}. This paper addresses the following question : how the dynamics of magnetic reconnection in an open-spine null-point topology can lead to inject energetic particles in the interplanetary medium over a large longitudinal range? Through the invoked mechanism, the angular spreading can lead particles to eventually reach the Earth-connected field line, even if the outer spine will never be connected to it.

\subsection{The Role of the Magnetic Reconnection Dynamics}

One first clue to answer this question can be obtained from the analysis of a former 3D interchange reconnection model driven by flux emergence \cite{RosGals10}. Particle acceleration and transport was there modeled by following test-particles in a fixed magnetic field taken from one snapshot of an MHD simulation.  They showed that particles are accelerated at the reconnection current sheet, and are injected along a long narrow layer of open field lines. The spatial extension of this propagation channel could be a way to get energetic particles in a large range of longitudes. One can still ask why such an extended acceleration region occurs in this simulation, and whether or not it is sufficient to explain the large spreading of energetic particle beams. According to \inlinecite{Aulanier_al07} and \inlinecite{Masson_al09b} the injection channels are affected by the dynamics of magnetic reconnection. Therefore, the time evolution of the magnetic field could play an important role in the extension of particle beams.

So as to address all these questions, we performed a new 3D MHD simulation of a coronal null-point topology with an open outer spine, driven by slow photospheric shearing motions, and we investigate the dynamics of the interchange reconnection. The model is not dimensionalized, but the topology being considered can make it relevant to a pseudo-streamer around an active region, or to a small jet inside a coronal hole. After describing the numerical model in Section~\ref{s-model}, we present the results on the magnetic reconnection dynamics in Sections~\ref{s-current} and \ref{s-slip}. Assuming that the dynamics of the magnetic field can represent the macroscopic evolution of injection and propagation channels, our results provide a new conceptual model to understand how energetic particles could reach the Earth-connected interplanetary field line when the reconnection site is initially located far away from it (Figure~\ref{f-fig1}). We finally discuss the relevance of our model on the interplanetary injection of flare-accelerated particles (Section~\ref{s-disc}).

\section{Model Description} 
\label{s-model} 

The initial magnetic configuration is potential and is built with sub-photospheric monopoles placed such as to obtain a bipole embedded in an open diverging magnetic field, leadinan asymmetric null-point topology with an open outer spine. The configuration is therefore well-suited to model large-scale pseudo-streamers above active regions, or small-scale jets inside coronal holes. The magnetic field is normalized so that the maximum initial photospheric magnetic field $B[z=0,t=0]= 1$. The top left panel on Figure~\ref{f-fig1a} displays the initial magnetic topology where grey field lines materialize the null-point separatrices.

We use the explicit 3D MHD code described in \inlinecite{AulDemGra05}, to which we added gravity finite-$\beta$ effects. The numerical simulation is performed in non-dimensionalized units and in a Cartesian domain in which $x$ and $y$ are the horizontal axes and $z$ the vertical axis, covering the volume defined by: $x\in[-37,23]$, $y\in[-30,30]$ and $z\in[0,100]$. The mesh is non-uniform with $201 \times 181 \times 201$ grid points, not counting two layers of ghost cells around each face of the domain. The mesh size results from a combination of geometric series in all directions and is adjusted to be the thinnest around the null point position ($z=4.81$) and at the inversion line ($z=0$). The mesh is displayed on Figure~\ref{f-fig1a} : on the top right panel, the black grid represents the cells of the mesh at the photospheric boundary, whereas the lower part ($z=[0,12]$) of the vertical mesh, where the fan-spine separatrices are located, is shown on the bottom panel with the white grid. (The details of the mesh can be seen by zooming in the electronic version of this figure.)

Our calculation solves the following set of equations using a fourth-order finite-difference and third-order predictor-corrector scheme:

\begin{equation}
\label{eqcont}
 \frac{\partial \rho}{\partial t} = - {\bf \nabla} \cdot (\rho {\bf u}) 
 + \kappa\, \Delta (\rho-\rho(t=0)) ,\nonumber 
\end{equation}

\begin{equation}
\label{eqmom}
 \frac{\partial {\bf u}}{\partial t} = - ({\bf u} \cdot
          {\bf \nabla}) {\bf u} + \rho^{-1}\, {\bf \jmath} \times {\bf B} 
          - \rho^{-1}\, {\bf \nabla} P + {\bf g}
          + \nu^\prime\, \mathfrak{D} {\bf u} ,\nonumber 
\end{equation}

\begin{equation}
\label{eqtp}
 \frac{\partial T}{\partial t} = - ({\bf u} \cdot {\bf \nabla}) T 
 - (\gamma-1)\, T\, {\bf \nabla} \cdot {\bf u}
 + \kappa\, \Delta (T-T(t=0)) ,\nonumber 
\end{equation}

\begin{equation}
\label{eqinduc}
 \frac{\partial {\bf B}}{\partial t} = {\bf \nabla} \times
         ({\bf u} \times {\bf B}) + \eta\, \Delta {\bf B} ,\nonumber 
\end{equation}

\begin{equation}
\label{eqamppres}
  {\bf \jmath} = {\bf \nabla} \times {\bf B} \, \, \,  , \, \, \, P = 2 \rho T \nonumber 
\end{equation}

\noindent $\rho$ being the mass density, $T$ the plasma temperature, ${\bf u}$ the plasma velocity, ${\bf B}$ the magnetic field. $\gamma$ is the ratio of specific heats chosen here at $1$. These equations are written in the code in their non-conservative fully developed form. The lower boundary of the numerical box follow the line-tied conditions, whereas the upper boundary and the edges of the box follow the open condition.

The operators $\nu^\prime \mathfrak{D} {\bf u} $ and $\eta \Delta {\bf B}$ correspond respectively to a viscous filter that uses a pseudo-Laplacian adapted to the local mesh \cite{AulDemGra05}, and a standard collisional uniform resistive term. The values of $\nu^\prime$ and $\eta$ lead to diffusive speeds such as:  $u_\nu= 7\%~c_{\rm A}^{\rm max}$  and $u_{\eta}= 5\%~ c_{\rm A}^{\rm max} $ on the scale of the smallest mesh, where the maximum Alfv\'en speed is $c_{\rm A}^{\rm max}=7$. $\kappa\, \Delta (\rho-\rho(t=0))$ and $ \kappa\, \Delta (T-T(t=0))$ are non-physical explicit diffusive terms which smooth the gradients of density and temperature at a speed of $u_\kappa =  7.5\%~ c_{\rm A}^{\rm max}$, and help to stabilize the numerical computation.

Owing to the high-order spatial and temporal scheme of our 3D MHD code, firstly the numerical diffusion is very small, and secondly current sheets always remain resolved on more than one grid point. Indeed, the use of the explicit resistive term is mandatory to stabilize the code (see the Appendix of \inlinecite{Aulanier_al05}). Also, the use of five grid points to calculate spatial derivatives imposes that all the gradients developing throughout the simulation are resolved on no less than three grid points (see the bottom panel of Figure~\ref{f-fig1a}). The explicit resistivity in our code is therefore responsible for all non-ideal processes described in this paper.

In order to drive magnetic reconnection, we prescribed a slow photospheric flow which moves the positive polarity toward the negative $y$ in the area of $x\in[-7,3]$ and $y\in[10,-30]$, where a part of the fan is rooted. The top right panel on Figure~\ref{f-fig1a} displays the location at the photospheric boundary and the shape of the applied shearing motion. A temporal ramp is applied, allowing an initial relaxation phase between $t=0$ and $t=15$. The photospheric driving is then constant until $t=40$, after which it is progressively stopped with a similar temporal ramp. The maximal amplitude of the prescribed photospheric velocity is: $u_{\rm phot}^{\rm max}= 0.35 \approx 5 \% ~c_{\rm A}^{\rm max}$.

Considering the large size of the numerical box and the existence of a null point, a finite-$\beta$ stratified atmosphere is required so as to respect the following coronal regime~: $\beta<1$ (except in a radius of 1 around the null point, where $\beta> 1$) and a sub-Alfv\'enic and sub-sonic regime in the whole domain for the coronal perturbations, directly induced by the shearing motions. 

We define this atmosphere as follows. We first fix the photospheric density $\rho[z=0,t=0]$ in order to obtain a maximal photospheric Alfv\'en speed $ c_{\rm A}^{\rm max}=7$. This characteristic speed defines the time unit required for an Alfv\'en wave to travel the distance between the centers of the two polarities of the small bipole, $D=7$.

We then define the gravity and temperature profiles such that the Alfv\'en speed remains roughly constant above the altitude of the null point. There, the density scale height, and therefore the pressure scale height $= 2T/g$ should be half of the magnetic scale height $=B/(\partial B / \partial z)$. Assuming an hydrostatic atmosphere, and choosing $2T/g$ as shown in Figure~\ref{f-fig2}, we integrate numerically the logarithm of the pressure and deduce the vertical density profile at $t=0$. Even though this atmosphere does not exactly reproduce solar numbers, it still puts the system in a solar-like regime.

\section{Null-Point Squashing and Magnetic Reconnection}
\label{s-current} 

\subsection{Current Sheet Formation}
 
The photospheric shearing motion of one part of the fan's footpoints leads to a bulging of field lines rooted below the fan surface. This expansion compresses the fan separatrix surface and the inner spine, leading to the development of an over-density (Figure~\ref{f-fig3}). 

This also leads to the shearing of the spine (bottom row, Figure~\ref{f-fig3}), {\it i.e.} a loss of alignment between the inner and the outer spine \cite{Pariat_al09}. This deformation of the null-point compresses and extends the region of weak $B$, inducing the formation of a narrow and intense current sheet in the vicinity of the null point and along the separatrices \cite{Low87,Aly90,lau93,RickardTitov96,Galsgaard_al03}, as seen on Figure~\ref{f-fig3}.  

Thus, null-point reconnection is expected to occur, leading to a realignment of the inner and the outer spine \cite{Antiochos_al02}. By plotting field lines from fixed footpoints and by integrating these field lines up to their conjugate footpoints at different times, we can follow the evolution of the connectivities during the simulation. 

\subsection{Interchange Reconnection}

In order to follow the field lines that are initially rooted below the fan surface, in the positive polarity advected by the applied flow, we computed analytically the position of their advected footpoints at each times. Therefore, we follow the connectivity of these field lines from their footpoints fixed in the applied photospheric flow. We initially plot three groups of field lines: pink, light blue and red, closed below the fan surface, and four groups of field lines (pink, light blue, yellow and red) open in the corona (bottom left panel of Figure~\ref{f-fig3}). 

In each group, field lines are plotted from fixed footpoints distributed along a small segment. All these segments have the same length and each of them is located at the same distance from the fan's footpoints. The field lines of the same color group are plotted from footpoints  uniformly distributed along one segment.
 As reconnection proceeds, magnetic flux is exchanged between connectivity domains, and the fan surface is displaced and sequentially crosses the fixed footpoint of the plotted field lines. Since the plotted field lines are uniformly distributed along each segment, the number of field lines that reconnected per time unit  is proportional to the reconnection rate.

As the system evolves, the field lines, initially closed in the inner connectivity domain, reconnect at the null point and jump to the outer connectivity domain, opening-up in the corona (bottom row of Figure~\ref{f-fig3}). On the opposite, the initial open field lines reconnect at the null point and closing-down below the fan surface. This exchange of connectivity between the inner and the outer connectivity domain is, in fact, the interchange reconnection \cite{Edmondson_al09}. During the reconnection through separatrices, the transfer of magnetic flux between the inner and the outer connectivity domain leads to displacements of the separatrices. Therefore, the positions of the fan surface and the spine line, both change. 

Even though the deformation of separatrices appears clearly during the evolution (see the dark blue field lines lines on Figure~\ref{f-fig3}) the relative length over which the spine field line shifts by reconnection, as compared to the horizontal scale over which the neighboring open field lines expand with height, is not very large. In itself, this result suggests that the large expansion with height of field lines above flaring active region, as identified by \cite{Klein_al08}, may not always allow energetic particles previously accelerated low down in the corona to reach the Earth-connected field lines. 

\section{Slip-Running Reconnection of Open Field Lines }
\label{s-slip}

\subsection{Slipping Field Lines} 

Before field lines reconnect at the null point, they display an apparent continuous slipping motion toward the inner spine (see Figure~\ref{f-fig3} as well as the animation on the electronic version). Figure~\ref{f-fig4} also displays the evolution of the connectivities at high altitude, in the whole numerical domain. The four panels show the same groups of field lines, as plotted in Figure~\ref{f-fig3}, at four different times. The animation shows that, right after having reconnected at the null point and jumped to the outer connectivity domain, field lines are located all along the outer spine in the whole numerical domain. Nevertheless, these reconnected field lines do not stay in the vicinity of the outer spine: they very quickly move farther away from it. Indeed, after the null point reconnection, field lines display an apparent slipping motion along a specific direction towards the negative $y$. 

Even though the topology is that of a null point, this apparent slipping motion is reminiscent of a slipping reconnection regime occurring through quasi-separatrix layers \cite{Aulanier_al06}. A quasi-separatrix layer (QSL) is a purely three dimensional geometrical object, formed by narrow volumes of continuous connectivity but across which the connectivity displays strong gradients  \cite{PriestDemoulin95}. Contrary to the null-point reconnection, the absence of magnetic discontinuity implies that, in a QSL, field lines simply exchange their connectivity with their neighbors and that their footpoints is moved in time across the QSL. Thus, continuous reconnection  between neighboring field lines leads to a continuous exchange of connectivity and apparent slippage of field lines plotted from fixed footpoints can be noticed. This slipping reconnection regime has been confirmed by several MHD numerical simulations \cite{Pontin_al05b,Aulanier_al06} and supports numerous association of observed flare ribbons with QSL calculated in extrapolated magnetic fields \cite{Demoulin_al97,Mandrini_al97}. 

Since it has recently been shown that this regime can also occur around null points with completely closed magnetic separatrices \cite{Masson_al09b,Torok_al09}, we investigate whether or not the same happens here, in the presence and around open separatrices. The slipping motion of reconnected field lines, before and after the null point reconnection, could result either from slip-running/slipping reconnection through QSLs, or from mass motions induced by the propagation of shear Alfv\'en waves injected either by the photospheric driving, by the reconnection jets, or even from numerical errors.

Note that at late times, the animation shows that the green field lines, rooted in the photospheric driving region  far from the null point, also move at the top of the numerical domain (Figure.~\ref{f-fig4}). Nevertheless, these motions are straightforward to interpret. They clearly result from the upward propagation of undulations, that simply correspond to the propagation of a shearing Alfv\'en wave from the line-tied boundary, which is generated by the photospheric driving. 

\subsection{Relation Between Plasma and Field Line Velocities} 

We compared the slipping velocities of field lines and the plasma velocities in the vicinity of these reconnected field lines, at the top of the box. We selected the fifth reconnected field line in each of the 3 studied groups of field lines, and we determined for these three selected field lines their slipping velocity at the top of the box for each time interval of the simulation. Then, we evaluated the plasma velocity at the top of these three selected field lines. Since the mesh of the plotting is much tighter than the mesh of the 3D MHD code, the position of the top of the field lines is never located exactly on one of the mesh points where the MHD variables are integrated. Thus, we computed the mean value of the plasma velocity of the four mesh points surrounding the field lines position at the top of the box.  The resulting velocities are plotted in Figure~\ref{f-fig5}.

After the field lines have reconnected at the null point, their slipping velocity is greater than the Alfv\'en speed, $ c_{\rm A} \simeq 1.5$, and than the fast-mode speed $c_{\rm F} \simeq 1.7$ at the top of the box. This leads to a super-Alfv\'enic and super-sonic slipping motion. The slipping velocities of field lines later decrease to $\approx 0.2$ and become sub-Alfv\'enic and sub-sonic. Meanwhile, the plasma velocity at the top of the box is found to increase during the simulation, in agreement with the upward propagation of the shear Alfv\'en wave from the photospheric driving at the bottom of the box. Even though the slipping velocities decrease, they remain much faster than the plasma velocity until $t=50$ (see Figure~\ref{f-fig5}). After $t=50$, the temporal evolution of the slipping velocities becomes the same as the evolution of the plasma velocities and their values becomes comparable. Therefore, the displacement of field lines at the end of the simulation is indeed due to mass motion, {\it i.e.} ideally evolving field lines, such as the green field lines. 

At this stage, the analysis suggests that the slipping motions of the reconnected field lines, at least until $t=50$, are due to magnetic reconnection through a QSL, and at least not to mass motion. The evolution of the slipping velocity of the field lines, successively super-Alfv\'enic then sub-Alfv\'enic, corresponds to a transition from slip-running to slipping reconnection following null-point reconnection, as first put foward by \inlinecite{Masson_al09b}, but for fully closed magnetic fields.

\subsection{Can Numerical Errors Induce the Slipping Motions ?} 

If the slipping motion of the reconnected field lines was due to systematic numerical errors throughout the whole integration domain, whether coming from the MHD computation or from the field line integration code, the green field lines, which evolve through the upward propagating shear Alfv\'en wave, should display a slipping motion too. The animation attached to Figure~\ref{f-fig4} clearly shows, however, that the reconnected field lines slip long before the top of the undulating green lines start to slip, the latter only being due to the Alfv\'en wave passing through the top boundary. This difference indicates that there is no significant systematic numerical error. 

Still, one expects larger errors which could locally affect the reconnected field lines, since those pass through a current sheet, in which sharper magnetic gradients develop. Such errors can be estimated with the divergence of the magnetic field, since our MHD code does not enforce the solenoidal condition ${\bf \nabla}\cdot{\bf B}=0$. It should be noted that all the MHD equations are fully developed in the code, with all the $\bf{\nabla}\cdot {\bf B}$ terms being removed, so that these terms neither produce momentum nor induction.  In order to quantify the effect of non-zero $\bf{\nabla}\cdot {\bf B}$ on the slipping motions, we nevertheless  study the dynamics of the field lines with modified magnetic field ${\bf B}^\star$. At each time and position in the domain, every component of the modified magnetic field was computed by the following Taylor-like expansion where $d_i$ is the local size of the mesh interval in the direction $i$~: $B^\star_i=B_i + d_i\, \bf{\nabla}\cdot {\bf B}$, where $i=x, y, z$. Using this modified magnetic field, we re-conduct the same velocity analysis as described above, for the very same reconnected field lines. We find that at the top of the box,  the temporal evolution of the slipping velocity and the displacement of the ${\bf B}^\star$ field lines display the same temporal evolution than the ${\bf B}$ field lines. The dynamics of magnetic reconnection for ${\bf B}^\star$ is the same than ${\bf B}$ : the succession of null-point, slip-running and slipping reconnection regimes. However, one notices that reconnection process for ${\bf B}^\star$ starts twice Alfv\'en time later than for ${\bf B}$. In addition, the slipping velocity of field lines, at a given location at the upper intersection of field line with the boundary,  differs between   ${\bf B}^\star$  and  ${\bf B}$. Indeed, we evaluate that the slipping velocities with ${\bf B}^\star$ depart from those with  ${\bf B}$ by $35\%$ on average.This number is presumably due to the strong sensitivity of the integration of field lines within a strongly diverging flux tube such as a quasi-sepatrix layers (which is described below as the structure along which field lines are slipping), even when starting from a region where B has locally been slightly modified to B*

Even though our MHD code does not ensure the solenoidal condition, and in spite of the fact that the present calculation was ran in single precision, we thus find that the measurable numerical errors do not affect the large-scale dynamics of magnetic reconnection and the observed slipping of the reconnected field lines. 

\subsection{An Elongated QSL around the Open Spine}

In \inlinecite{Masson_al09b}, we had shown that the closed separatrices of a null point were surrounded by a QSL halo. The presence of a QSL, through which continuous magnetic reconnection is known to occur, can be checked by computing the norm $N$ \cite{PriestDemoulin95} or the squashing degree  $Q$ \cite{Titov_al02}. These quantities  somehow measure the gradients of field line connectivity, and thus they map the position of the QSL. 

Usually $N$ and $Q$ are calculated in closed field configurations only. Here we compute $Q$ in our open fields in a similar way as how \inlinecite{PriestDemoulin97} calculated $N$~: the upper and bottom boundaries are considered as the two reference planes in which field line footpoints are calculated. Even if the lower line-tied boundary appears as more natural reference plane than the upper open frontier at $z=100$, $Q$ is uniquely defined for each lines given this choice of computation limit. However, the choice of a different reference plane (such as a different altitude for the top boundary) would result in a different value of $Q$ for a given field line. Nevertheless, assuming a monotonous divergence of the magnetic field, the distribution of $Q$ would remain monotonously homogenous with a different choice of $z$. Hence, the occurrence and locations of QSLs in open field lines, manifested by open narrow layers of high-$Q$, would be independent of the choice for the reference upper plane.

Figure~\ref{f-fig6} shows the location and shape of the high-altitude QSL which surrounds the open spine, as seen from the large $Q$ isocontours plotted at $z=99.5$. Initially, the QSL is broad and is almost symmetrically distributed around the spine. At later times, however, the prescribed photospheric motions shear the system, and increase the asymmetry of the null point. Following \inlinecite{Parnell_al96}, the parameter $p$ which defines the asymmetry of the fan, increases in our simulation from $1.2$ (at $t=0$) to $2$ (at $t=65$). This is associated with the development of a non-symmetric, elongated and thin QSL around the open spine, in a similar way as what we found earlier in an initially asymmetric, but closed, null-point topology \cite{Masson_al09b}.

On Figure~\ref{f-fig5}, we report the distribution of the high value of Q, and thus the location of the QSL, at the top of the box. It is now clear that the intersections of the slipping field lines with the upper boundary of the box are located within the QSL. This finally leads us to interprete the super- (followed by sub-) Alfv\'enic field line motions as a substantial amount of slip-running (followed by slipping) reconnection, of the field lines which have previously jumped from the inner to the outer connectivity domain by the interchange null-point reconnection.

The distribution of Q (Figure~\ref{f-fig6}) can also explain why the slipping velocity of some reconnected field lines displays a local minimum around $t=40$ after the slip-running phase (see Figure~\ref{f-fig5}). A band of low Q has formed within a higher Q region. \inlinecite{Aulanier_al06} noted that field lines slip slower (res. faster) in regions of lower (resp. larger) $Q$ . We also find that the brief period of lower field line velocities around $t=40$ is indeed associated with the crossing of this local low-$Q$ region by the slipping field lines.

\section{Discussion}
\label{s-disc}

\subsection{MHD Results}

So as to explain several in-situ measurements in the heliosphere and at Earth, we
investigated an MHD model for the dynamics of 3D magnetic reconnection to address the interplanetary injection of energetic particles during impulsive-SEP events. The model is based on the interchange reconnection mode in an open-spine null-point topology \cite{Edmondson_al09,Pariat_al09}. This situation is relevant for the two solar magnetic configurations that couple the closed magnetic corona with the open corona and the heliosphere~: the large-scale pseudo-streamers above active regions, and the small-scale jets inside coronal holes. In a pseudo-streamer configuration, flare-related particles may have been previously accelerated inside the inner connectivity domain, whereas in a jet configuration, particles are supposed to be directly accelerated at the null-point. In both cases all energetic particles pass through the current sheet formed at the interface between open and closed field regions, before being injected in the interplanetary medium through interchange reconnection.

The main result of the MHD simulations is that, before and after the standard interchange reconnection at a coronal null point, the field lines rooted in the photosphere continuously reconnect through slip-running and slipping reconnection regimes. This is due to the presence of a halo of quasi-separatrix layers (QSLs) that surrounds the fan separatrix surface and spine singular field line originating at the asymmetric null point. In such an ``hybrid topology'' magnetic reconnection does not behave as usual, {\it i.e.} field lines do not exchange their connections instantaneously by pairs. This continuous reconnection associates with a large extension of the current sheet along the fan, exceeding the length of the current sheet which forms at the shearing null point. This large current sheet is a direct result of the development of a QSL around the spine line of a null point becoming increasingly asymmetric. This QSL sustains slip-running and slipping reconnection, hence leads to a large displacement of slipping field lines at high altitudes in the open field domain. This result generalizes to open spine configurations the same MHD process originally discussed for a fully closed magnetic fields \cite{Masson_al09b}. Since slipping reconnection across QSLs around an asymmetric (also called improper) null point seems to occur for line-tied cases as well as for open boundary conditions, we argue that it must be a generic property of  3D reconnection at null points. As such, the ``torsional fan reconnection mode'' found in asymmetric nulls by \inlinecite{HachamiPontin10} may actually be just another formulation of the same physical effect.

Even though our calculations do not treat the dynamics of the particles themselves, our analyses of the evolution of reconnected field lines can be used as a first proxy to determine the time-evolution of injection and propagation channels for accelerated particles. Thereby, we assume that the particle acceleration process, wherever it occurs, can lead to a significant number of particles at sufficiently high energies.

\subsection{Possible Consequences for Particles}

In the classical two-dimensional interchange reconnection mode, particles should only be injected into open field lines through the short current sheet that forms around the misaligned separatrices of the null point. Thus, particles should only be injected along the few field lines which have reconnected at the null point. The QSL-related field line dynamics in our simulation, however, let us conjecture that during 3D interchange reconnection, particles can also be successively injected along (initial fast) slip-running (and later slow
slipping) field lines.

	In principle, the association of slip-running  and interchange reconnection should allow that particles are accelerated all along the intense current sheet localized in the QSL and not only around the null point. Let us illustrate a discretized version of this continuous mechanism~: when an initially closed field line $(i)$ reconnects and opens at the null point, a first set of particles $(i)_1$ are injected along this reconnected open field line. Then, as the same open field line $(i)$ slip-runs away from the open spine, another closed field line $(i+1)$ reconnects and opens at the null, and a second set of particles $(i+1)_1$ are injected along this second reconnected field line. In the meantime, the motion of the field line $(i)$ has made it slip along the QSL-related low-altitude current sheet and a new set of particles$(i)_2$ is injected along the open field line $(i)$. As this whole process repeats itself $n$ times, newer and newer sets of particles are accelerated and injected along the field line $(i+n)$ being just reconnected at the null point, as well as along all the slip-running field lines $(i+n-1; i+n-2; ... ; i+1; i)$ that are moving farther and farther away from the open spine. 
Therefore, this mechanism implies that the set of energetic particles would be injected continuously along field lines which are undergoing slip-running/slipping reconnection away from the open spine.

We conclude that during the interchange followed by slip-running reconnection, particles could be injected within an extended and narrow flux tube, progressively formed by the field lines slipping along a large-scale open QSL. In a fixed magnetic field configuration, this may explain the elongated shape of particle impacts at the top of the numerical domain, as obtained in the simulations of \inlinecite{RosGals10}. In a time-varying system, our analysis implies that particle beams originating from the interchange reconnection should sweep the high altitude domain, along a narrow QSL-related flux tube that extends through the slipping reconnection process.

\subsection{ Adaptation to the Heliospheric Context}

\begin{figure}
\centerline{
\includegraphics[width=1\textwidth]{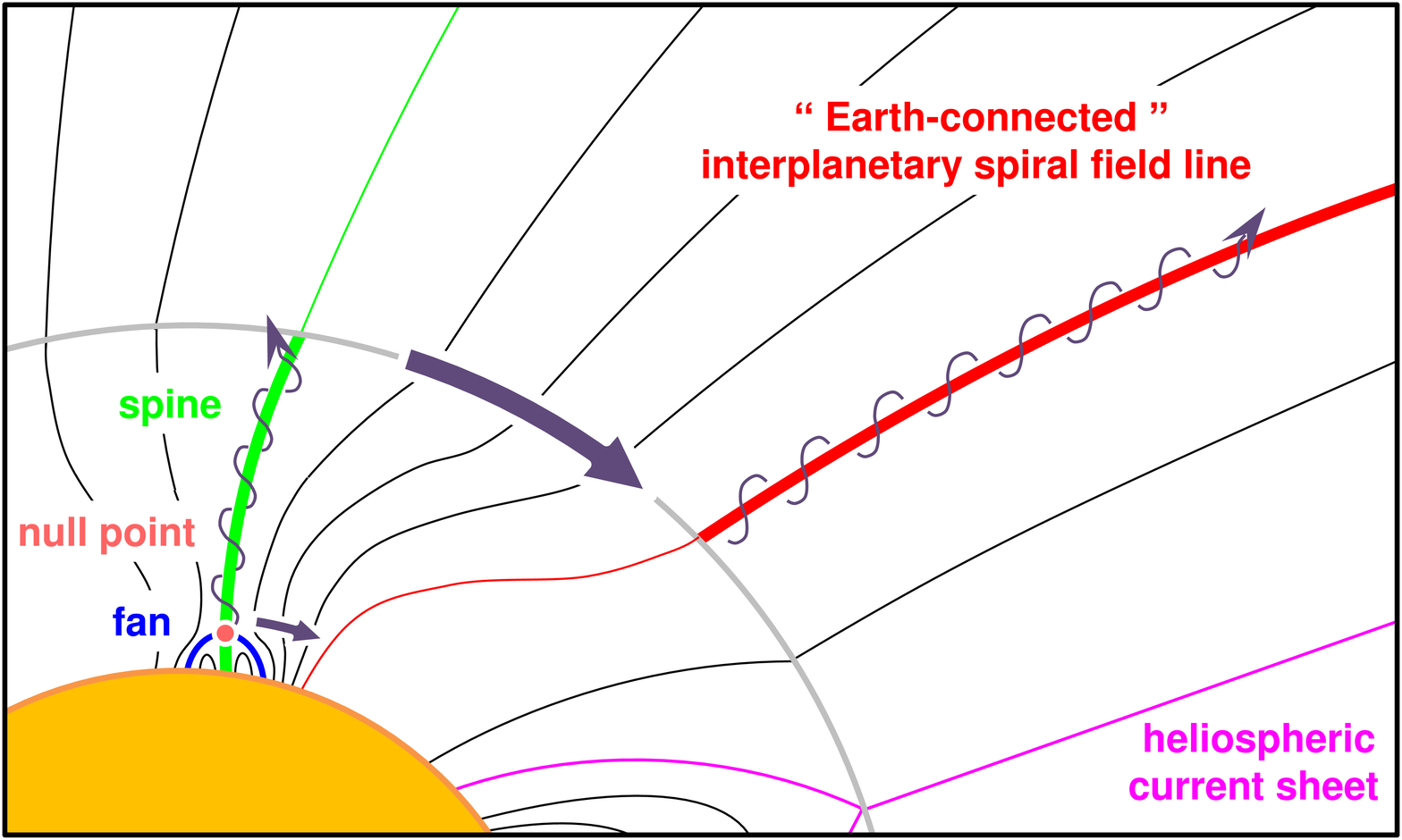}
}
\caption{Cartoon displaying the connectivity problem between the open coronal field lines from the acceleration site (green line) and the Earth-connected parker spiral field line (red line). 
 }
\label{f-fig1}
\end{figure}

\begin{figure}
\centerline{
\includegraphics[width=1\textwidth,bb= 118 100 531 650,clip=]{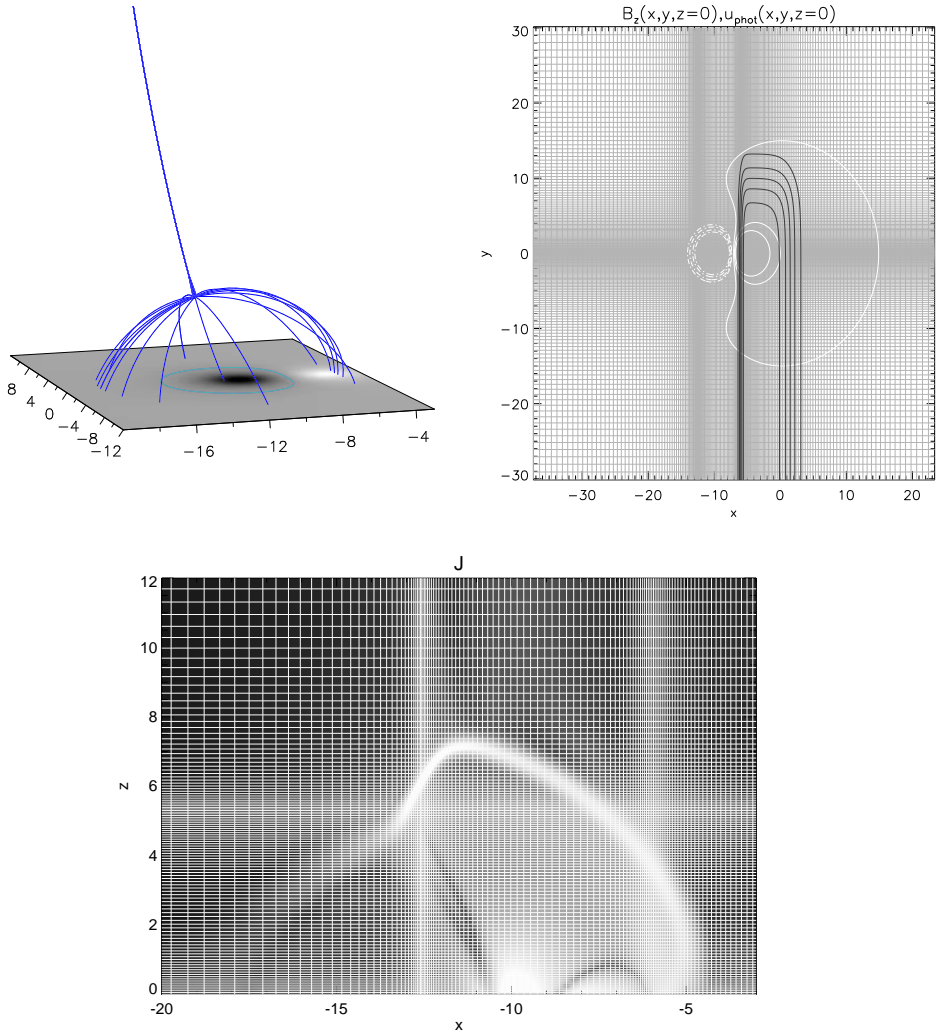}
}
\caption{Top left panel: Initial magnetic topology displaying the separatrix field lines in grey. At the photospheric boundary, the greyscale shading represents the vertical magnetic field $B_z$ at the photosphere, $z=0$. The blue field lines forming the dome materialize the fan surface and the singular field line intersecting perpendicularly the fan at the null point corresponds to the spine. Top right panel : Photospheric distribution of the mesh. The vertical magnetic field and the velocity flow in the ($x,y$) plan.  The grey grid displays the non-uniform photospheric mesh,  the white dashed contours are the negative isocontours of the vertical magnetic field for $B_z= -0.75, -0.5, -0.25, -0.1$, at $z=0$, and the white solid contours represents the positive isocontours of the vertical magnetic field for $B_z=0.5, 0.75, 1$ at  $z=0$. The black contours displays the photospheric velocity field, orientated toward the negative-y. Bottom panel : Vertical two dimensional ($x,z$) cut of the density distribution $|j|$, respectively on the left and on the right panel, at $t=50$ and at the y position of the null point ($y=0$). The white grid correspond to the mesh in the ($x,z$) plan at $y=0$. (A full-resolution version of this figure is available in the online version of the paper) }
\label{f-fig1a}
\end{figure}

\begin{figure}
 \centerline{
  \includegraphics[width=1\textwidth,bb=22 173 579 707, clip=]{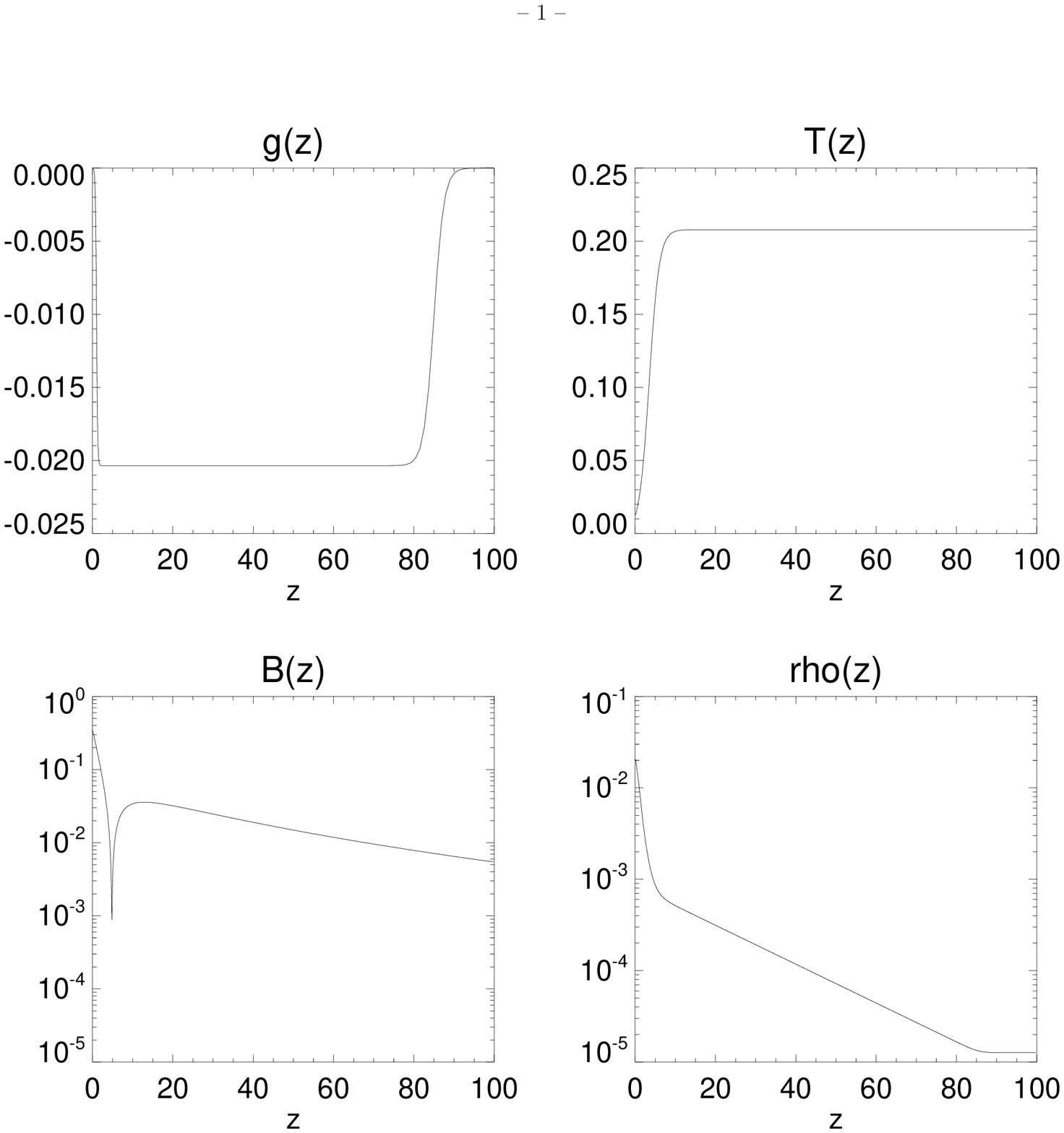}
}
\caption{Vertical profiles of the gravity (top left), the temperature (top right), the magnetic field (bottom left) and the density (bottom right) at the null point position. }   
\label{f-fig2}
\end{figure}

\begin{figure}
 \centerline{
 \includegraphics[width=1\textwidth,bb=35 222 575 687,clip=]{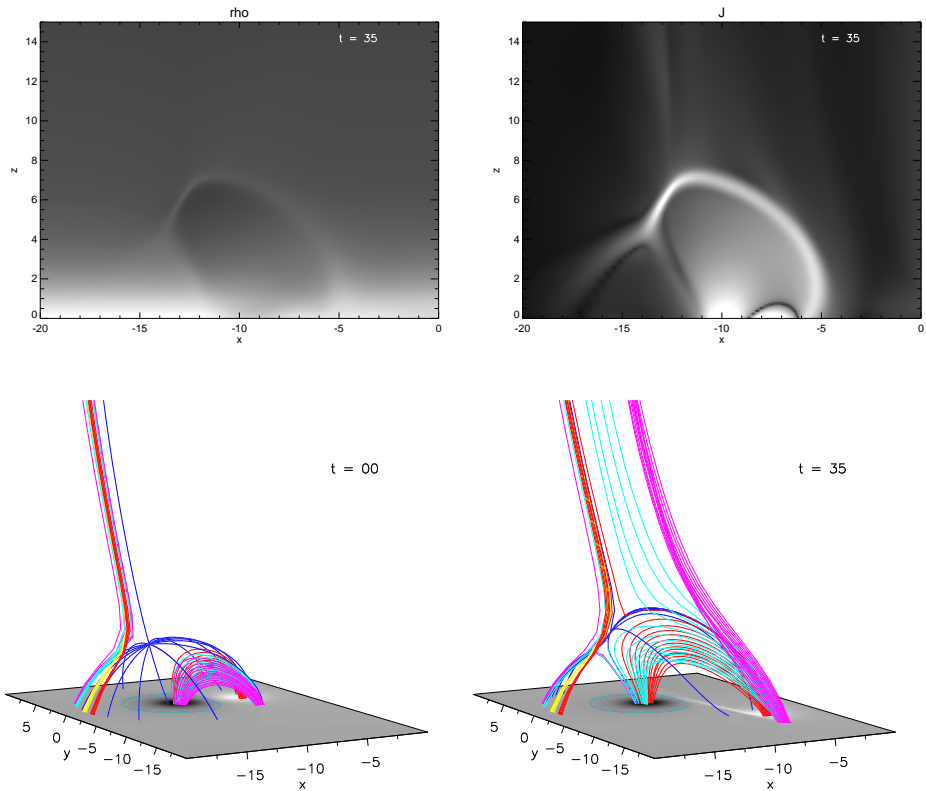}
}
\caption{(top row) Vertical two dimensionnal ($x,z$) cut of the current density $\rho$ and of the density distribution $|j|$, respectively on the left and on the right panel, at $t=35$ and at the y position of the null point ($y=0$). (bottom row) The left and the right panel display respectively the null point configuration at the initial time and at $t=35$. The dark blue lines represent the fan and the spine separatrices passing through the null point. The colored field lines initially located in the outer connectivity domain are plotted from fixed footpoints and the group of colored field lines initially located below the fan surface are plotted from footpoints which are fixed in the advected flow. The distribution of the photospheric vertical magnetic field $B_z(z=0)$ is coded by a greyscale.
}   
\label{f-fig3}
\end{figure}

\begin{figure}
\centerline{
 \includegraphics[width=1\textwidth,bb=87 26 523 700,clip=]{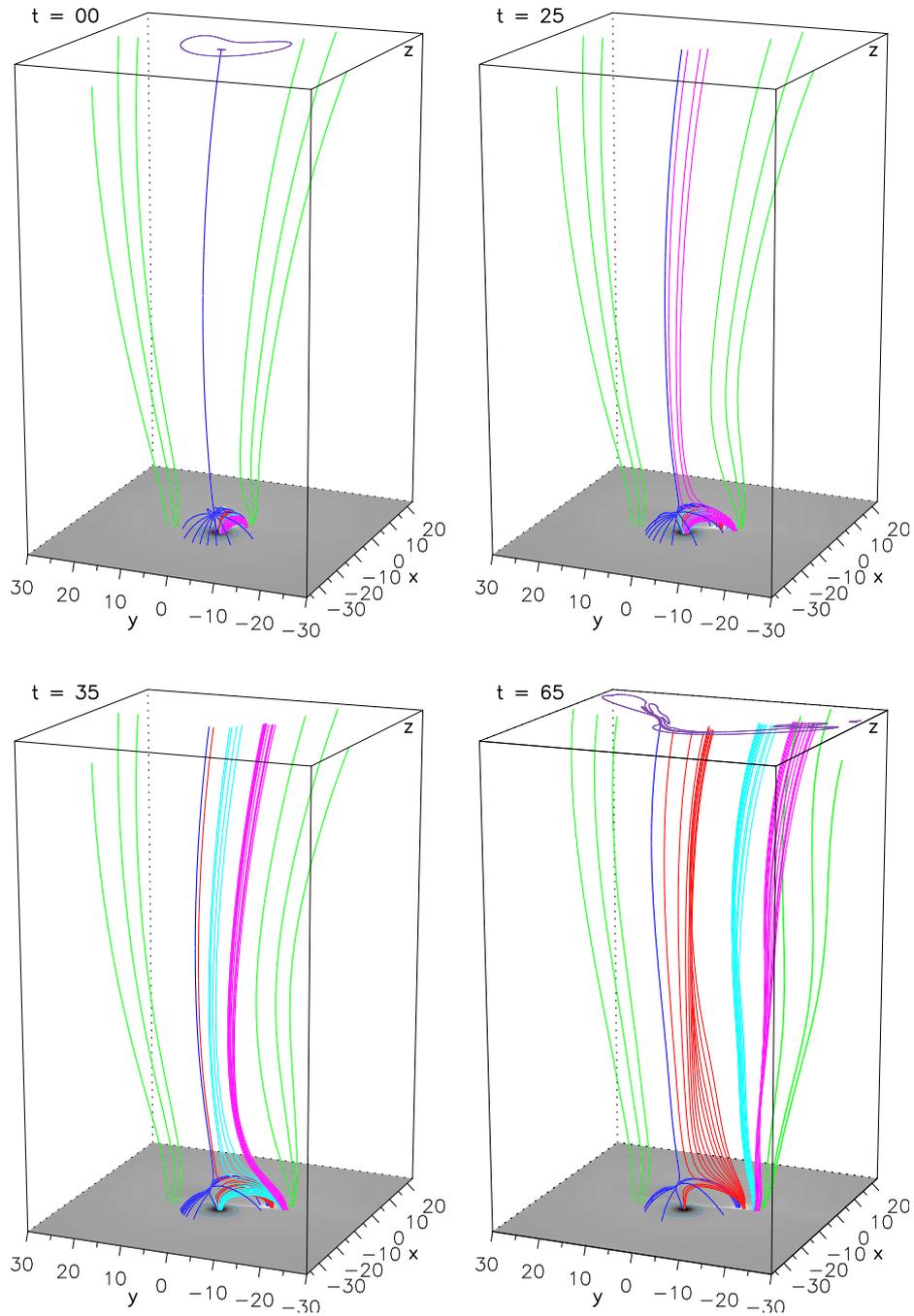}
}
\caption{Evolution of selected field lines. The dark blue field lines are the spine and fan field lines passing through the null point. The green field lines represent the open diverging magnetic field. The other colored field lines are plotted with fixed footpoints in the prescribed photospheric flow.The distribution of the vertical photospheric magnetic field $B_z(z=0)$ is coded in grey scale. Panel (a) : initial ($t=0$) magnetic configuration : all colored field lines are located below the fan surface in the inner connectivity domain. Panels (b), (c) and (d) : evolution of magnetic field lines in the outer connectivity domain. On the first and fourth panels, the purple isocontours plotted at the top of the box corresponds to the spatial distribution of the logarithm of squashing factor for $\log Q=1.12, 4.01$ and $6.90$. (An animation is available in the electronic version of the journal)
} 
\label{f-fig4}
\end{figure}

\begin{figure}
\centerline{
\includegraphics[width=1\textwidth]{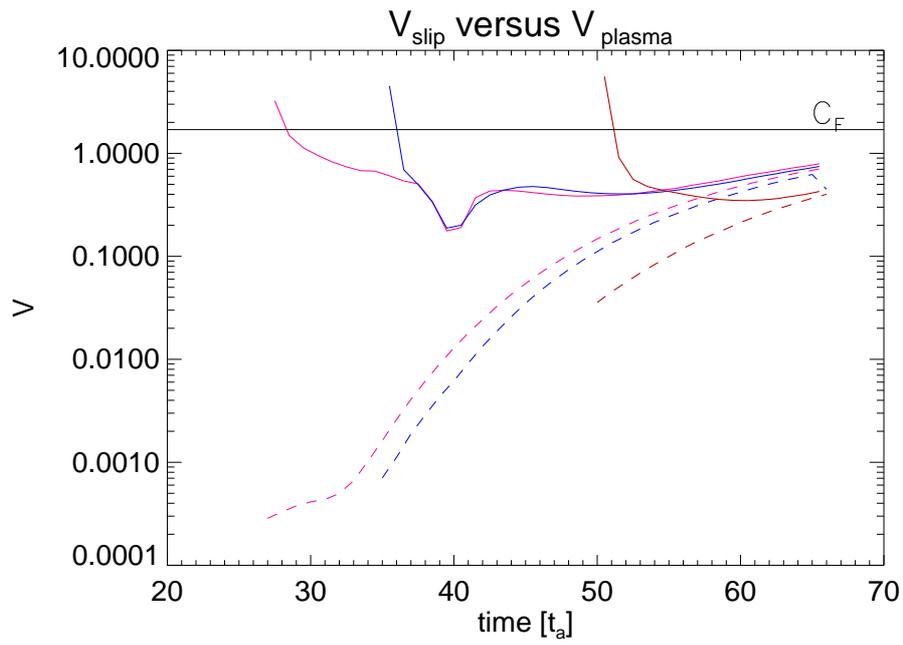}
}
\caption{Evolution of the slipping (solid lines) and plasma (dashed lines) velocity of the three selected field lines. The color of the curves corresponds to the color of the selected field line. The horizontal line corresponds to the fast-mode speed, $c_{\rm F}=\sqrt{c_A^2+c_S^2}$, at the top of the box}
\label{f-fig5}
\end{figure}

\begin{figure}
\centerline{
\includegraphics[width=1\textwidth,bb=0 150 550 780,clip=]{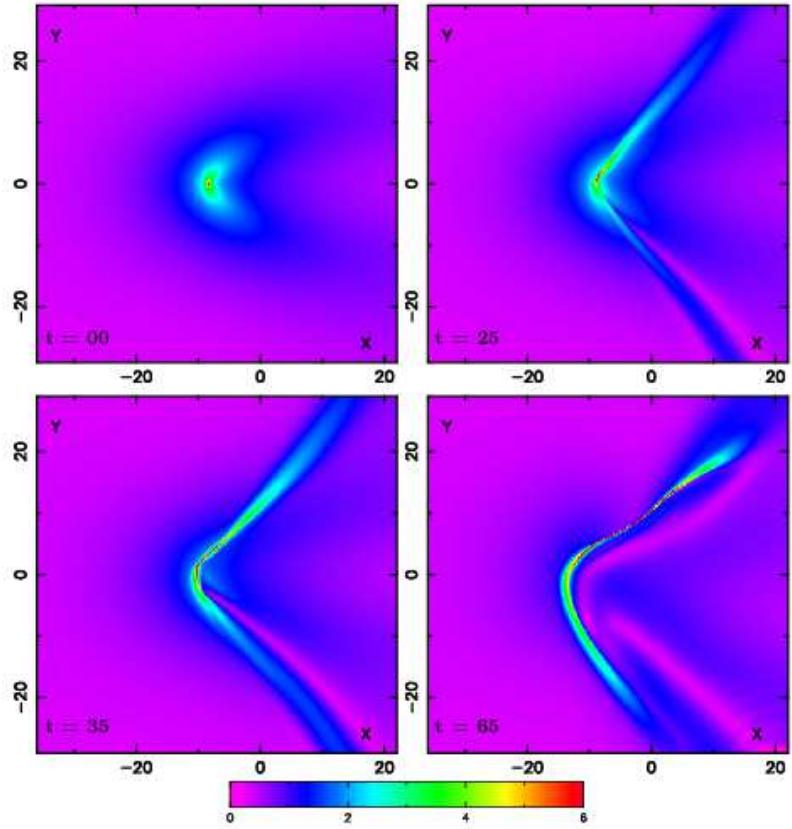}
}
\caption{Temporal evolution of the QSL mapping at the top of the numerical box, $z=99.5$, representing by the squashing factor $\log Q$ in color scale in the range $1-6$. The four panels correspond to the same time of Figure~\ref{f-fig4}.}
\label{f-fig6}
\end{figure}

When considering a solar-like configuration such as a large-scale pseudo-streamer or a small-scale jet , the resulting sweeping energetic particle beam could eventually reach the Earth-connected Parker spiral field line, even if the acceleration site is not initially connected to this very line (as shown in Figure~\ref{f-fig1}).  Our model provides a new type of MHD-based particle injection process along a sweeping beam, that may cover a large angular range.

Around $80~\%$ of active regions related to impulsive SEP events are close to or included in an open coronal magnetic flux tube \cite{Wang_al06,Pick_al06}. They are auspicious magnetic configurations for finding pseudo-streamer and jet configuration. In addition, in such a region nearby open coronal magnetic field, when we measured a SEP events at the Earth, a small jet is usually observed in UV and occurs at a comparable time than type III bursts  \cite{Wang_al06,Nitta_al06}.This suggests that open and close field lines are reconnecting through the interchange mode in a null-point topology. Therefore, the null-point configuration embedded in an open magnetic field, as shown in Figure~\ref{f-fig1}, can be considered as a typical configuration for flare-related impulsive SEP events, regardless any CME. This leads us to suggest  that the model proposed by this study should be a common injection process. However, one can ask if the angular distance travelled by slipping field lines and thus the sweeping particle beam, in a solar context, is significant ?

Even though the non-dimensionalized units do not allow to directly compare the results of the simulation with the observations, one can estimate the angular distance, $\theta$, covered by slipping field lines for typical solar values. Associating the geometry of the system to a right triangle, where the spine and the top of the box form the right angle and the slipping field line is the hypothenuse, we evaluate the angle $\theta$. For an Alfv\'en speed of $c_{\rm A}\approx 500~\rm{km\;s^{-1}}$, and an arbitrary distance unit of $D=10~\rm{Mm}$ for a coronal jet (resp. $100~\rm{Mm}$ for a pseudo-streamer), the flux tube formed by the slipping field lines expands over an angular range  $\theta \simeq 14~^{\circ}$ at the altitude of $142\rm{Mm}$ for a jet (resp. $2~\rm{R_\odot}$). Comparing t the spine field line's motion covering the angular distance $\theta_{\rm spine}\simeq 1.4~^{\circ}$, it appears that the slipping reconnection lead to a much more substantial angular spreading of the open coronal flux tube, guiding energetic particles in the heliopshere.

In addition, it is noteworthy that $\theta$ is constrained by the simple geometry of the the model used in the 3D-MHD simulation. Indeed, the diverging of the magnetic field in 3D is relatively small, but a largest expansion will lead to increase values of $\theta$. Actually, the spatial extension of the open coronal magnetic field being super-radial \cite{Wang_al06,Klein_al08}, field lines could easily slip over larger distances than in our simulation, implying that the particle beam could sweep over a larger angular range.

According to \inlinecite{Droege10} and \inlinecite{CholletGiacalone11}, energetic particles should not be scattered significantly during their interplanetary propagation, implying that energetic particles propagate in the inner heliosphere along the same field line.  Thus, the association of slipping and interchange reconnection provides a new way to understand SEP connection to the Earth and the large angular range over which particles can be detected in the interplanetary medium, suggesting that the cross-field transport of accelerated particles in the coronal magnetic field is not the only way to inject particles in an extended flux tube. This result leads us to believe that the MHD model proposed in this study can be fully relevant to explain several observations such as the measurement of energetic particles over more than $90 ^{\circ}$ of longitude.

The model that we propose for sweeping particle beams in the heliosphere is based on the assumption that the macroscopic spatial evolution of propagation channels follows the dynamics of interchanging and slip-running field lines.
  Future work will therefore have to calculate explicitly the particle trajectories in the diffusion region
and along the continuously reconnecting field lines in a time-dependent magnetic field.

%
\acknowledgements
The authors thank the referee for helpful comments, which improved the clarity of the paper.
The MHD calculations were done on the quadri-core bi-Xeon computers of the 
Cluster of the Division Informatique de l'Observatoire de Paris (DIO). 
The work of S.M. is funded by a fellowship of Direction G\'en\'erale 
de l'Armement (DGA).
Financial support by the European Commission through the FP6 SOLAIRE Network 
(MTRN-CT-2006-035484) is gratefully acknowledged. 
%
%

%


%
%

%
%

%
%

%
%

%

%
%

%
%
%
%
%
%
\end{article}
\end{document}